# Influence of random opinion change in complex networks

Yi Yu
School of Electrical and Electronic Engineering,
Nanyang Technological University,
Singapore.
e-mail: yyu6@e.ntu.edu.sg

Gaoxi Xiao
School of Electrical and Electronic Engineering,
Nanyang Technological University,
Singapore.
e-mail: egxxiao@ntu.edu.sg

*Abstract*—Opinion formation in the population has attracted extensive research interest. Various models have been introduced and studied, including the ones with individuals' free will allowing them to change their opinions. Such models, however, have not taken into account the fact that individuals with different opinions may have different levels of loyalty, and consequently, different probabilities of changing their opinions. In this work, we study on how the non-uniform distribution of the opinion changing probability may affect the final state of opinion distribution. By simulating a few different cases with different symmetric and asymmetric non-uniform patterns of opinion changing probabilities, we demonstrate the significant effects that the different loyalty levels of different opinions have on the final state of the opinion distribution.

## I. INTRODUCTION

Opinions tend to be highly diversified in modern societies and the diversified opinions strongly affect almost every single aspect of our life, from as big as election [1, 2], nation-level or region-level policies, to as "small" as interpersonal relationship [3] and purchasing decisions, etc. There have been existing studies on the propagation of different opinions in social networks [4-6] and the impacts of opinion propagation on the social structures [7, 8], etc. An equally important topic is how people make consensus with each other, which is known as the opinion formation problem.

Extensive studies have been carried out on the opinion formation problem and quite a few models have been proposed [9-19]. Among them include the voter model [9-11], which assumes that a randomly chosen individual (or a node in the network; hereafter "individuals" and "nodes" shall be used interchangeably) may adopt the opinion of his randomly selected neighbor; and bounded confidence model [12-14], assuming that the opinion of a randomly chosen individual may be the weighted average of all those neighbors whose opinions are not too much different from his own one. Another closely related popular model is the Deffuant model [15], which assumes that each time a randomly chosen node may interact with only one of its randomly selected neighbor. More specifically, an opinion tolerance $d$ is introduced: if the difference between the opinions held by the two nodes is less than $d$, they may make consensus to further reduce the difference between their opinions; otherwise nothing would happen. Study results show that such a model would lead to a final state of opinion distribution with several groups, each of which holding its own local consensus. With a decreasing value of opinion tolerance $d$, the number of opinion groups increases and transits at certain set of values.

Random change of individual's opinion was introduced into the Deffuant model for the first time in [17]. Specifically, a node may randomly change its opinion at a certain rate. With the random change of opinion being introduced into the model, instead of reaching local consensus in the final state, opinions will have a well-defined distribution with several bell-curve-style peaks. Follow-up work has been conducted to evaluate the impacts of initial state on the final opinion distribution [18]. Extensive simulation results show that, except for some extreme cases (e.g., the initial state is to have a single global consensus), different initial states essentially lead to the same final opinion distribution.

An important issue, however, has been largely missed in these studies to the best of our knowledge. That is, how the different levels of "loyalty" to different opinions affect the final state of the opinion distribution. Our daily experiences may easily tell us that people holding certain opinions tend to change their opinions more easily or more difficultly than other opinion holders. An example may be the conversion between different religions [20].

In this paper, we study on how different changing probabilities of different opinions may affect the final opinion distribution in the system. Specifically, we shall still adopt the Deffuant model to perform the simulations of opinion formation. However, instead of using a uniform distribution of opinion changing probability, we consider a few non-uniform distribution cases where individuals with different opinions have different probabilities to randomly change their mind. With some simulation results, our preliminary study reveals the strong impacts of non-uniform loyalty-level distribution on the final opinion distribution.

The rest of the paper is organized as follows: section 2 briefly introduces the Deffuant model with non-uniform random opinion changes. Simulation results and discussions are presented in Section 3. Section 4 concludes the paper.

## II. MODEL DESCRIPTION

### A. Review of the Deffuant model with random opnion change

Deffuant model [15] assumes that time is slotted into time steps, and opinions are continuously and uniformly distributed in the interval [0, 1] at the beginning. At each time step $t$, a randomly chosen node $A$ and its randomly chosen neighbor $B$ carry out the operation as follows. Denote their opinions at time $t$ as $o(t, A)$ and $o(t, B)$ respectively. If the difference between $o(t, A)$ and $o(t, B)$ is less than $d$, the two nodes shall make consensus following the rule below:

$$o(t+1, A) = o(t, A) - \mu[o(t, A) - o(t, B)]; \quad (1)$$
$$o(t+1, B) = o(t, B) + \mu[o(t, A) - o(t, B)].$$

A smaller value of $\mu$ may slow down the evolution process while it is believed that different values of $\mu$, as long as it is within the range (0, ½], eventually lead to the same steady state [15]. Hereafter we let $\mu = 1/2$ for simplicity, as that in most of the existing studies. The work of [17] first introduced *noise* to simulate the free will of individuals. Specifically, it assumes that at each time step, a randomly chosen node has a certain probability $p$ to randomly change its opinion. Hereafter we term such random change of opinion as *mutation*. To highlight that individuals holding different opinions may have different mutation probabilities, we introduce a mutation probability function $P(x)$ to reveal the probability that an individual holding opinion $x$ may change its opinion in each time step.

### B. Non-uniform opinion change

All the existing studies, to the best of our knowledge, have assumed a uniform distribution of mutation probability for different opinions, that is $P(x) = p$ for all $x$. In this work, we study on the case where $P(x)$ could be of a non-uniform distribution.

We consider several different symmetric and asymmetric mutation probability functions. For the asymmetric functions, we consider the simplest case where the mutation probability increases or decreases linearly with the opinion. Specifically, the class of asymmetric mutation probability function is defined as:

$$P(x) = \alpha(x - 0.5) + p, \quad (2)$$

This allows us to examine the effects of different slopes of the function.

For the class of symmetric functions, we examine the simple one where the mutation probability increases or decreases linearly with the distance to the central opinion. Specifically, we define the function as follows:

$$P(x) = \begin{cases} \alpha(x - 0.25) + p, & 0 \le x \le 0.5; \\ -\alpha(x - 0.75) + p, & 0.5 < x \le 1. \end{cases} \quad (3)$$

For the asymmetric cases we set the slope value as $\alpha = -0.02, -0.01, 0.01, 0.02$; and for the symmetric cases we let $\alpha = -0.04, -0.02, 0.02, 0.04$. Note that for all these cases, the average mutation probability remains the same as being equal to $p$. In the rest of the paper, we let $p = 0.01$. Figure 1 illustrates $P(x)$ for all the eight cases that we are going to examine in this paper.

## III. INFLUECNE OF MUTATION PROBABILITY

We apply different mutation probability functions $P(x)$ on the Deffuant model and observe the influences in two aspects: *i)* the change in bifurcation pattern, referring to the appearance, splitting and merging of peaks in the opinion distribution as we change the opinion tolerance $d$; and *ii)* the change in the final opinion distribution.

### A. Bifurcation patterns

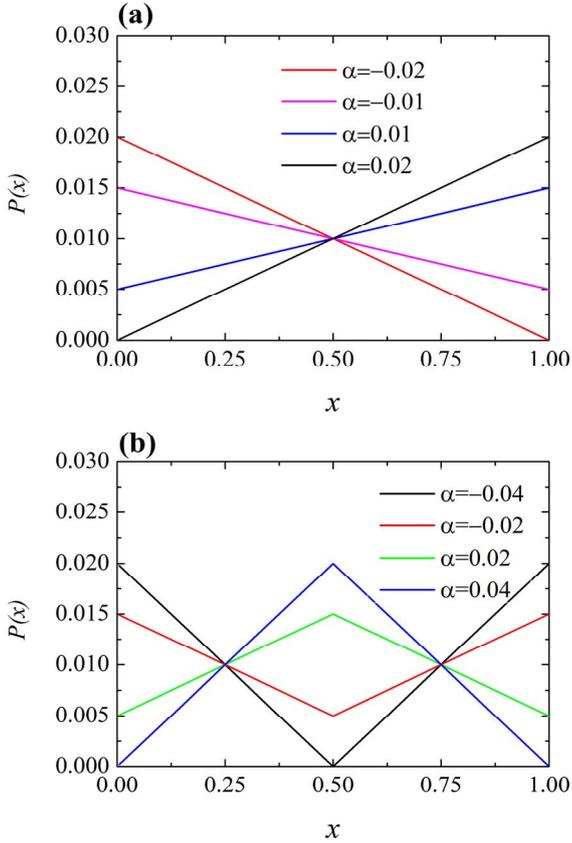

Figure 1. Different mutation probability functions $P(x)$ used in this paper: (a) asymmetric cases; (b) symmetric cases.

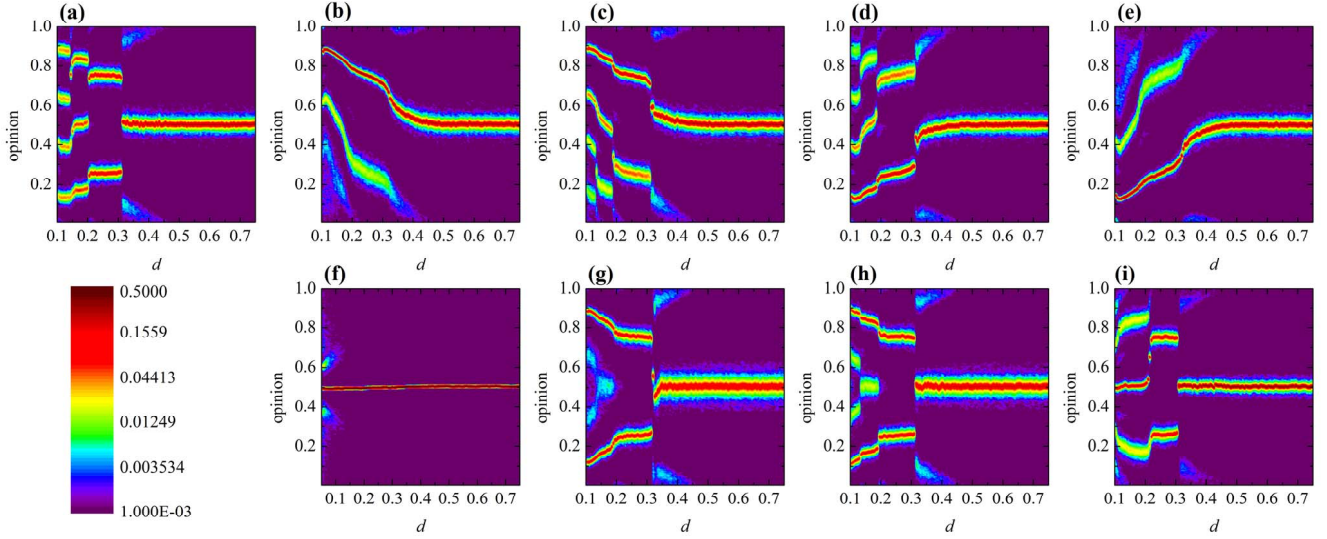

Figure 2 Bifurcation pattern for (a) uniform *P(x)*; asymmetric *P(x)* with (b) α=-0.02; (c) α=-0.01; (d) α=0.01 (e) α=0.02, and symmetric *P(x)* with (f) α=-0.04; (g) α=-0.02; (h) α=0.02; and (i) α=0.04.

An important observation on the Deffuant model is its bifurcation pattern when system evolves. When we decrease the opinion tolerance $d$, transition of peak numbers would happen. In the existing results which assume $P(x) = p$ for all opinion $x$, the number of peaks $n$ roughly follows $n \sim 1/2d$ [16]. In this part, we shall observe how the bifurcation pattern changes with different mutation probability functions $P(x)$.

We carry out the simulation on the ER random network [21] with a size of $N = 10^4$ and an average degree of 10. For each $P(x)$, we generate 10 networks and for each network, we increase the value of $d$ from 0.1 to 0.75 with a step length of 0.005. For every $d$ value, we perform the simulation for $t = 5 \times 10^7$ time steps, long enough for the network to reach the final steady state. For each round of simulation, we average the opinion distribution of the last 1000 steps as the final state opinion distribution.

Figure 2 shows for different uniform and non-uniform functions $P(x)$, the evolution of the bifurcation pattern with different values of $d$. The color map illustrates the average in 10 networks. In all cases, we observe peaks in opinion distribution which are shown as red strips in the figures. Figure 2(a) is for the case with the uniform mutation probability function where $P(x) = p$. Figures 2(b) to 2(e) show the steady-state opinion distribution when we adopt the asymmetric mutation probability function as defined in Equation (2) with $\alpha = -0.02, -0.01, 0.01, 0.02$ respectively; and Figures 2(f) to 2(i) are for the cases with symmetric mutation probability functions as defined in Equation (3) with $\alpha = -0.04, -0.02, 0.02, 0.04$ respectively.

Figure 2(a) clearly illustrates the transitions in the number of peaks with a changling value of $d$; between transitions, the steady-state distribution of opinion stays almost unchanged. This is consistent with the observations from previous studies [14-18]. Looking at Figures 2(b) to 2(e) where $P(x)$ are asymmetric functions, we find a combined gradual and sudden changes in the steady-state opinion distribution as $d$ changes, and for most time, the change is gradual and continuous. This becomes more obvious when $d$ is of a relatively larger absolute value: as we could observe in Figures 2(b) and 2(e), the steady-state opinion distribution basically changes continuously with $d$. Taking Figure 2(e) as an example, we see that as $d$ increases, the locations of the peaks continuously shift towards larger values, and the peaks originally at high opinion values for small values of $d$ slowly go diminished. For the cases with symmetric functions $P(x)$, the opinion distribution also changes with $d$, yet in a symmetric pattern, and the evolution of the bifurcation pattern is quite different for different cases where the mutation rate at central opinion is of the highest or lowest value respectively. Take Figure 2(g) as an example where the opinions closer to central value have relatively lower mutation probabilities. We observe that as $d$ increases, the locations of the peaks shift towards the central value. The two peaks closest to the center firstly merge into a single peak and gradually diminish. Meanwhile, the outside peaks continue to move towards the center, until finally merge into a single peak at the center after a transitional change when $d$ is large enough. For the case where opinions closer to central value have realtively higher mutation probabilities, we may take Figure 2(i) as an example. We can observe when $d$ increases from 0.1 to 0.2, the locations of peaks gradually shift away from the center. The trend of moving away from the center continues though it becomes weaker when $d$ gets larger to be from 0.2 to 0.3. Finally, when $d$ is large enough, a single peak emerges at the center.

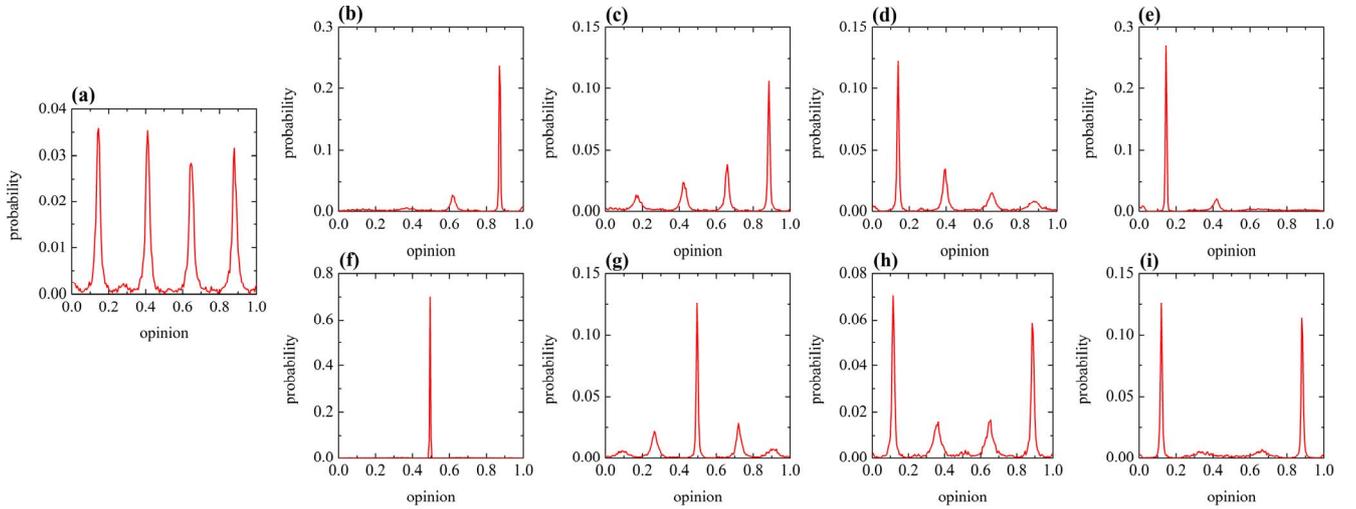

Figure 3 Steady-state opinion distribution for (a) uniform $P(x)$, asymmetric $P(x)$ with (b) α=-0.02; (c) α=-0.01; (d) α=0.01 (e) α=-0.02, and symmetric $P(x)$ with (f) α=-0.04; (g) α=-0.02; (h) α=0.02; and (i) α=0.04.

Combining the above discussions, the overall observation is that, for the cases with non-uniform mutation probability functions, the opinion distribution has a combination of gradual and sudden changes with the change of $d$. Generally speaking, when the opinion tolerance $d$ increases, opinion peaks located at opinions with higher mutation probabilities would vanish first and peaks located at opinions with lower mutation probabilities would gradually shift towards the location of the just-disappeared peaks. Such observations are very interesting and request detailed further studies in the future.

*B. Opinion distribtion*

We now have a closer look at the steady-state opinion distribution. Figure 3 shows the scope of steady-state opinion distribution where $d = 0.1$ for the 9 cases which are arranged in the same order as those in Figure 2. In Figure 3(a) where the mutation probability is uniform for all opinions, the several peaks have approximately the same height. Such observations are consistent with those in the existing studies [15-19]. For the 8 non-uniform cases, peaks would generally appear to be of a lower height at opinion values with relatively higher mutation probabilities. Comparing Figure 3(b) and Figure 3(c), we see that with greater value of $|\alpha|$, which indicates a more heterogeneous mutation probability distribution for different opinions, the heights of different peaks also appears to be more heterogeneous.

Overall, the results reported in this section may well resemble our daily observations: an opinion enjoying a higher level of loyalty among its followers tends to have a larger number of followers; and opinion groups with less tolerance to opinion change may also tend to grow up their size as long as the tolerance is not going too low to scare away potential followers (or equivalently, to lower the chance that individuals holding other opinions may change to hold this opinion).

An interesting observation we should point out is that, as that in previous studies on cases with uniform mutation function, the steady-state opinion distribution under non-uniform mutation function is also largely independent of the initial opinion distribution with the exception of only some very special cases (Detailed discussions on the independence of the final steady state to the initial state however have to be omitted in this paper due to length limit.). Such may open the new possibility to observe and estimate the loyalty levels of different opinions in a society by observing their evolution, and their (relatively) steady state if applicable.

IV. CONCLUSION

In this paper, we studied the influences of non-uinform opinion changes on the formation of different opinions in complex networks. Specifically, we adopted the Deffuant model with random opinion change. By simulating a few different cases where different opinion holders may have different probabilities to change their mind, we demonstrated the corresponding changes in the bifurcation pattern and the steady-state opinion distribution. Overall, the observation is that peaks located at the opinions with higher mutation probabilities tend to have lower values.

Future studies will be carried out on (i) possibility and methods (if applicable) for estimating relative mutation rates of different opinions by observing the evolution and the state of different opinions; and (ii) evaluating the cases where target of opinion change is not randomly distributed, in other words, where different opinions have different levels of attractiveness to holders of other opinions (which arguably may be related to the opinion's tolerance towards "change of mind").

ACKNOWLEDGMENT

This work is partially supported by Ministry of Education (MOE), Singapore, under research grant RG 28/14 and MOE2013-T2-2-006.

REFERENCES

[1] B. R. Berelson, P. F. Lazarsfeld, and W. N. McPhee, "Voting: a study of opinion formation in a presidential election," Chicago: University of Chicago Press, 1954.

[2] B. Norrander, "Measuring state public opinion with the senate national election study," State Polit. Policy Q., vol. 1, pp.111-125, 2001.

[3] S. L. Parker, G. R. Parker, and J. A. McCann, "Opinion taking within friendship networks," Am. J. Polit. Sci., vol. 52, pp. 412-420, 2008.

[4] A. M. Timpanaro and C. P. C. Prado, "Generalized Sznajd model for opinion propagation," Phys. Rev. E, vol. 80, pp. 021119, 2009.

[5] F. Amblard and G. Deffuant, "The role of network topology on extremism propagation with the relative agreement opinion dynamics," Physica A, vol. 343, pp. 725-738, 2004.

[6] Y. Wang, G. Xiao, and J. Liu, "Dynamics of competing ideas in complex social networks," New J. Phys., vol. 14, pp. 013015, 2012.

[7] F. Wu and B. A. Huberman, "Social structure and opinion formation, arXiv:cond-mat/0407252v3, 2004.

[8] R. S. Burt, "The Social Capital of Opinion Leaders," The Ann. Am. Acad. Polit. Soc. Sci., vol. 566, pp. 37-54, 1999.

[9] R. A. Holley and T. M. Liggett, "Ergodic theorems for weakly interacting infinite systems and the voter mode", Ann. Prob., vol. 3, pp. 643-663, 1975.

[10] V. Sood and S. Redner, "Voter model on heterogeneous graphs," Phys. Rev. Lett., vol. 94, pp. 178701, 2005.

[11] C. Castellano, D. Vilone, and A. Vespignani, "Incomplete ordering of the voter model on small-world networks," EPL-Europhys. Lett., vol. 63, pp. 153, 2003.

[12] R. Hegselmann and U. Krause, "Opinion dynamics and bounded confidence:models, analysis and simulation," JASSS-J. Artif. Soc. S., vol. 5, 2002.

[13] J. A. N. Lorenz, "Continuous opinion dynamics under bounded confidence: a survey," Int. J. Mod. Phys. C, vol. 18, pp. 1819-1838, 2007.

[14] M. Pineda, R. Toral, and E. Hernández-García, "Diffusing opinions in bounded confidence processes," Eur. Phys. J. D, vol. 62, pp. 109-117, 2011.

[15] G. Deffuant, D. Neau, F. Amblard, and G. Weisbuch, "Mixing beliefs among interacting agents," Adv. Complex Syst., vol. 03, pp. 87-98, 2000.

[16] E. Ben-Naim, P. L. Krapivsky, and S. Redner, "Bifurcations and patterns in compromise processes," Physica D, vol. 183, pp. 190-204, 2003.

[17] M. Pineda, R. Toral and E. Hernández-García, "Noisy continuous-opinion dynamics," J. Stat. Mech. Theory E., vol. 2009, pp. 08001, 2009.

[18] A. Carro, R. Toral and M. S. Miguel, "The role of noise and initial conditions in the asymptotic solution of a bounded confidence, continuous-opinion model," J. Stat. Phys., vol. 151, pp. 131-149, 2013.

[19] M. Pineda, R. Toral and E. Hernández-García, "The noisy Hegselmann-Krause model for opinion dynamics," Euro. Phys. J. B, vol. 86, pp. 1-10, 2013.

[20] D. A. Snow and R. Machalek, "The sociology of conversion," Ann. Rev. Soc., vol. 10, pp. 167-190, 1984.

[21] P. Erdos and A. Renyi, "On random graph I," Publ. Math. Debrecen, vol. 6, pp. 290, 1959.